\documentstyle[11pt]{article}
\begin{document}
\parindent 0.7cm

\begin{titlepage}

\centerline{\large\bf
Two Time-dependent Solutions }
\centerline{\large\bf
of Magnetic field Annihilation in Two Dimensions}
\vspace{1cm}

\centerline{Xue-Shang Feng, Yong Liu, Feng-Si Wei and Zhan-Yin Ye }
\vspace{0.5cm}
\centerline{Laboratory of Numeric Study for Heliospheric Physics (LHP)}
\centerline{Chinese Academy of Sciences, P. O. Box 8701,
Beijing 100080, P. R. CHINA}

\vspace{2cm}

\centerline{\bf Abstract}
\vspace{0.6cm}

In this paper, two classes of exact analytic time-dependent
soultion of magnetic annihilation for incompressible magnetic
fluid, have been obtained by solving the magnetohydrodynamic (MHD)
equations directly. The solutions derived here possess scaling
property with time $t$ as the scale factor. Based on these two
solutions, we find that, for some given inflow fields, the
evolution of the annihilating magnetic field can be described by
the solutions of certain ordinary differential equations whose
variables are dilated simply by time $t$. The relevant evolution
characteristics in the process of magnetic annihilation are also
revealed.
\\\\
PACS number(s): 95.30.Q, 52.30, 75.50.M

\end{titlepage}

As is well known, magnetic reconnection is a very important
physical phenomena in many astrophysical objects. It is believed
that, magnetic reconnection can serve as such a mechanism to
transfer the magnetic energy into heat and kinetic energy of the
plasma \cite{sweet1958,parker1963}.

Generally, due to the nonlinearity of the MHD equations, it is very
difficult to find its analytic solutions. We have to turn to the
numerical simulations. But, unfortunately, numerical simulations
are usually limited by artificial resistivity and the
implementation of sensible boundary conditions
\cite{biskamp1994,forbes1987}.

Anyway, the exact analytic solutions are needed and helpful. On one
hand, they can provide the important insights into the problem
concerned, and on the other hand, they can be used to test the
relevant numerical methods. Just based on the beautiful exact
solutions, Parker found that, diffusion is an essential part of the
reconnection of lines of force at a neutral point \cite{
parker1973}. Since then, a lot of excellent analytic solutions have
been found and applied to analyze some concrete models
\cite{sonnerup1975, phan1990,jardine1992,heyn1993,heyn1996}.

Recently, Watson, Craig and Henton etc. have made a great progress
in searching for the exact solutions
\cite{craig1995,craig1996,watson1996,watson1997}. They found that,
exact families of reconnection solutions can be constructed in both
two and three dimensions, and, these solutions display the
essential characteristics of fast reconnection. Furthermore, they
have shown that, the reconnection solutions can be obtained by
superposing the velocity and magnetic fields of simple magnetic
annihilation model.

So, annihilation model can provide a solid basis for our
understanding the reconnection process. Up to now, there are few
works on the time-dependent case \cite{heyn1996,biernat1987}. Most
of the previous works in searching for the exact solutions
concerntrate on the case of the steady MHD. But, it is desired to
remove the restrictions of stationarity and incompressiblety etc.
for the realistic application.

In this paper, we present two time-dependent analytic solutions of
the incompressible MHD. We begin by introducing the basic equations
in section 1. The solutions are obtained in section 2. And, the
relevant physics indicated by the solutions is analysized in
section 3. Finally, the conclusions are summarized in section 4.

\vspace{0.5cm}
\centerline{\bf 1. The Basic Equations}

The incompressible MHD equations can be written as the following
nondimensional form \cite{watson1996}:
\begin{eqnarray}
\label{first}
\frac{\partial \vec{v}}{\partial t} &=&
-(\vec{v} \cdot \nabla) \vec{v} -\nabla P + (\nabla \times \vec{B})
\times
\vec{B}\\
\label{second}
\frac{\partial \vec{B}}{\partial t}&=&
\nabla \times (\vec{v} \times \vec{B}) + \eta \nabla^2 \vec{B}\\
\label{third}
\nabla \cdot \vec{B} &=& 0 \\
\label{fourth}
\nabla \cdot \vec{v} &=& 0
\end{eqnarray}
where the length, magnetic field $\vec{B}$, velocity $\vec{v}$,
time $t$, gas pressure $P$ and the resistivity $\eta$ have been
non-dimensionalized by the typical coronal parameters -
length-scale $L_c$, background field strength $B_c$, Alfven speed
$v_A=B_c/(8 \pi \rho)^{1/2}$, Alfven travel time $\tau_A=L_c/v_A$,
magnetic pressure $B_c^2/(8 \pi)$ and $v_A L_c$ respectively.

By taking the curl of Eq.(\ref{first}), we can remove $P$ in above
equations and get
\begin{equation}
\label{fifth}
\frac{\partial (\nabla \times \vec{v})}{\partial t}=
-\nabla \times [(\vec{v} \cdot \nabla) \vec{v}]
 + \nabla \times [(\nabla \times \vec{B})
\times \vec{B}]
\end{equation}
Now, we need to deal with Eqs.(\ref{second}, \ref{third},
\ref{fourth}, \ref{fifth}). Once we solve them and get $\vec{v}$
and $\vec{B}$, we can calculate the pressure $P$ according to
Eq.(\ref{first}).

\vspace{0.5cm}
\centerline{\bf 2. Magnetic Annihilation Solution}

In order to solve Eqs.(\ref{second}, \ref{third}, \ref{fourth},
\ref{fifth}), suppose that $\vec{B}, \vec{v}$ have the following
forms
\begin{eqnarray}
\label{sixth}
\vec{B}&=&\{\; 0,\;\; Y(x, t),\;\; 0 \;\}\\
\label{seventh}
\vec{v}&=&\{\; U(x, t),\;\; y V(x, t),\;\; 0\; \}
\end{eqnarray}
for the annihilation solution. Here, $U, V, Y$ are the functions of
$x$ and $t$. And thus, Eq.(\ref{third}) is satisfied automatically.

Now, Eqs.(\ref{fourth}, \ref{fifth}) and Eq.(\ref{second}) give
\begin{eqnarray}
\label{eighth}
\frac{\partial}{\partial x}U(x, t)+V(x, t)&=&0\\
\label{ninth}
\frac{\partial}{\partial t}V(x, t)+ U(x, t)\frac{\partial}{\partial x}
 V(x, t)+V(x, t)^2+f(t)&=&0\\
\label{tenth}
\frac{\partial}{\partial t}Y(x, t)+\frac{\partial}{\partial x}
(U(x, t) Y(x, t))- \eta \frac{\partial^2}{\partial x^2}Y(x, t)&=&0
\end{eqnarray}
where, $f(t)$ is a function of $t$ to be determined later by the
boundary condition.

If we further suppose that
$$ U(x, t)=U(x) g(t)$$
from Eq.(\ref{eighth}), it is evident that
$$ V(x, t)=V(x) g(t).$$
Substitute the above two equations into Eqs.(\ref{eighth},
\ref{ninth}), we achieve
\begin{eqnarray}
\label{1th}
\frac{\partial}{\partial t}g(t)+ c_1 g(t)^2 &=&0\\
\label{2th}
f(t)+c_2 g(t)^2&=&0\\
\label{3th}
\frac{\partial}{\partial x}U(x)+V(x)&=&0\\
\label{4th}
U(x) \frac{\partial}{\partial x}V(x)+V(x)^2-c_1 V(x)-c_2&=&0
\end{eqnarray}
with $c_i\; (i=1, 2, 3,...)$ being the (integral) constants. Now,
we are in position to state how to solve $g(t), U(x)$ and $V(x)$
from the above equations.

Eq.(\ref{1th}) tells us that
\begin{equation}\label{5th}
g(t)=\frac{1}{c_1 t + c_3}.
\end{equation}
Starting out from Eq.(\ref{3th}, \ref{4th}), we can obtain two
solutions.

A. Take $U(x)$ as the linear function of $x$, then we have
\begin{eqnarray}
\label{6th}
U(x) &=& c_4 x+c_5\\
\label{7th}
V(x)&=& -c_4\\
\label{8th}
c_2&=&c_1 c_4+c_4^2.
\end{eqnarray}

B. Let $U(x)=P(x)+c_4$, and substitute it into Eqs.(\ref{3th},
\ref{4th}), we arrive at
\begin{eqnarray}
\label{9th}
\frac{\partial}{\partial x}P(x)+V(x)&=&0\\
\label{10th}
P(x) \frac{\partial}{\partial x}V(x)+V(x)^2+c_4
\frac{\partial}{\partial x}V(x) -c_1 V(x)-c_2&=&0.
\end{eqnarray}
if we further have $P(x)=c_5 V(x)$ and $\frac{\partial}{\partial
x}V(x)=c_6 V(x)$, Eqs.(\ref{9th}, \ref{10th}) can be satisfied
easily. Then, we get
\begin{eqnarray}
\label{a1th}
V(x)&=& c_7 e^{c_6 x}\\
\label{a2th}
U(x)&=& c_5 c_7 e^{c_6 x}+c_4\\
\label{a3th}
c_6 c_4&=& c_1\\
\label{a4th}
c_5 c_6&=& -1\\
\label{a5th}
c_2&=& 0.
\end{eqnarray}

In the next step, inserting these solutions into Eq.(\ref{tenth}),
we can finally solve the magnetic field.

To give a feeling on the form of the magnetic field $\vec{B}$, let
us consider the asymptotic solution of $Y(x, t)$. When $\eta
\rightarrow 0$, Eq.(\ref{tenth}) is simplified as
\begin{equation}\label{b4th}
\frac{\partial}{\partial t}Y(x, t)+\frac{\partial}{\partial x}
(U(x, t) Y(x, t))=0.
\end{equation}
Take Eqs.(\ref{5th}, \ref{6th}) and Eq.(\ref{a2th}) into account,
we get
\begin{equation}\label{b5th}
Y(x, t)=c_9 \frac{(c_1 t + c_3)^{\frac{c_8}{c_1}}}{(c_4
x+c_5)^{1+\frac{c_8}{c_4}}}
\end{equation}
for solution A, and
\begin{equation}\label{b6th}
Y(x, t)=c_9 e^{-c_6 x} (c_4 e^{-c_6 x}+c_5 c_7)^{\frac{c_8}{c_4
c_6}-1} (c_1 t+c_3)^{\frac{c_8}{c_1}}
\end{equation}
for solution B.

With a suitable choice of the relevant constants, we can obtain
\begin{eqnarray}
\label{b7th}
Y(x, t)&=& \frac{1}{x^3 (t+1)^2}\\
\label{b8th}
Y(x, t)&=& \frac{e^x}{(e^x-1)^3 (t+1)^2}
\end{eqnarray}
corresponding to A and B. Here, the singularity exists evidently at
the neutral interface $x=0$ for both A and B. The presence of the
singularity is naturally unavoidable when diffusion vanishes, as
discussed by Parker \cite{parker1973}.

Eq.(\ref{tenth}) tells us taht, $Y(x, t)$ can not be simply
decomposed as the form of $X(x)T(t)$, after the similar supposition
for $U(x, t)$ is made. To solve the magnetic field $\vec{B}$
analytically, a certain transformation is needed.

For solution {\bf A}, $U(x, t)=\frac{c_4 x+c_5}{c_1 t+c_3}$.
Suppose $Y(x, t)=Y(s)$, and
\begin{equation}\label{x1}
s=(c_4 x+c_5)(c_1 t+c_3)^{\alpha}
\end{equation}
with $\alpha=-\frac{1}{2}$, Eq.(\ref{tenth}) can be transformed
into
\begin{equation}\label{x2}
\eta \frac{\partial^2}{\partial s^2}Y(s)+\frac{c_1-2 c_4}{2 c_4^2}
s \frac{\partial}{\partial s} Y(s)- \frac{1}{c_4} Y(s)=0.
\end{equation}
and for solution {\bf B}, $U(x, t)=\frac{c_5 c_7 e^{c_6 x}+c_4}{c_1
t+c_3}$, similarly, let
\begin{equation}\label{x3}
s=c_5 c_7 e^{c_6 x}(c_1 t+c_3)^{\alpha}
\end{equation}
with $\alpha=-1$, Eq.(\ref{tenth}) can be read as
\begin{equation}\label{x4}
c_6 \eta s \frac{\partial^2}{\partial s^2}Y(s)+(\eta c_6-s
)\frac{\partial}{\partial s} Y(s)-Y(s)=0.
\end{equation}

Now, the partial differential equation Eq.(\ref{tenth}) has been
transformed into the ordinary differential equations, whose
solutions can be expressed as special functions. The solutions of
Eq.(\ref{x2}) and Eq.(\ref{x4}) are
\begin{equation}\label{x5}
Y(s)= C_1 {\bf {}_1F_1}[-\frac{c_4}{c_1-2 c_4}, \frac{1}{2},
\frac{(-c_1+2 c_4)s^2}{4 c_4^2 \eta}] + s C_2 {\bf {}_1F_1}
[\frac{1}{2}-\frac{c_4}{c_1-2 c_4}, \frac{3}{2}, \frac{(-c_1+2
c_4)s^2}{4 c_4^2 \eta}]
\end{equation}
and
\begin{equation}\label{x6}
Y(s)=C_1 e^{\frac{s}{c_6 \eta}}+C_2 e^{\frac{s}{c_6 \eta}} {\bf
\Gamma}[0, \frac{s}{c_6 \eta}]
\end{equation}
respectively. Here, $C_1, C_2$ are the integral constants, ${\bf
{}_1F_1}$ is the Kummer confluent hypergeometric function while
{\bf $\Gamma$} the incomplete gamma function.

Up to now, we have obtained two classes of the exact analytic
solutions for both velocity field and magnetic field. We would like
to give a short comment here. It is deserved to note that,
Eqs.(\ref{x1}, \ref{x3}) are actually the scale transformation with
time $t$ as the scaling factor. So, our solutions will perform
certain scaling characteristic, i.e., when the velocity fields are
taken as Eqs.(\ref{6th}, \ref{a2th}, \ref{5th}), the evolution of
the magnetic field with time decribed by the solution of
Eq.(\ref{tenth}), can be described by the solution of the ordinary
differential equation Eqs.(\ref{x2}, \ref{x4}) dilated simply by
time $t$ according to $(c_1 t+c_3)^{\alpha}$.

\vspace{0.5cm}
\centerline{\bf 3. Physical Characteristics of the Solution }

In this section, we discuss the physics implied by the above
solutions. In doing so, we have to fix the free parameters. The
relevant functions are taken as
\begin{equation}\label{a6th}
g(t)=\frac{1}{t+1}
\end{equation}
together with
\begin{eqnarray}
\label{a7th}
U(x)&=& -x\\
\label{a8th}
V(x)&=& 1
\end{eqnarray}
for solution {\bf A}, and
\begin{eqnarray}
\label{a9th}
U(x)&=& e^{-x}-1\\
\label{a10th}
V(x)&=& e^{-x}
\end{eqnarray}
for solution {\bf B}. In fact, there is another possible choice
such as $U(x)=1-e^{x}, \; V(x)=e^{x}$ with g(t) given by
Eq.(\ref{a6th}) for the flow field of the annihilation solution,
but it is very similiar to the one given by
Eq.(\ref{a9th},\ref{a10th}), so we will not discuss it here.

The velocity field for solutions {\bf A} and {\bf B} are
\begin{equation}\label{b2th}
\vec{v}=\{\;-\frac{x}{t+1},\;\; \frac{y}{t+1},\;\; 0\; \}
\end{equation}
and
\begin{equation}\label{b3th}
\vec{v}=\{\;\frac{e^{-x}-1}{t+1},\;\; \frac{y e^{-x}}{t+1},\;\; 0 \;\}
\end{equation}
respectively. They are symmetric (for A) or asymmetric (for B)
stagnation-point flow weakening with time. For asymmetric case -
solution B, the flow field is shown in Fig.1.

With the given velocity fields $U(x, t)$, the equation of magnetic
field $B(x, t)$ can be read as
\begin{equation}\label{x7}
\eta \frac{\partial^2}{\partial s^2}Y(s)+\frac{3}{2}
s \frac{\partial}{\partial s} Y(s)+ Y(s)=0
\end{equation}
with the solution
\begin{equation}\label{x9}
Y(s)=-C s {\bf {}_1F_1} [\frac{5}{6}, \frac{3}{2}, -\frac{3 s^2}{4
\eta}] \;\;\;\;\;\; (\; s=-\frac{x}{\sqrt{t+1}} \;)
\end{equation}
and
\begin{equation}\label{x8}
\eta s \frac{\partial^2}{\partial s^2}Y(s)+(s + \eta
)\frac{\partial}{\partial s} Y(s)+Y(s)=0
\end{equation}
with the solution
\begin{equation}\label{x10}
Y(s)=C e^{-\frac{s}{\eta}}({\bf
\Gamma}[0, -\frac{s}{\eta}]- {\bf
\Gamma}[0, -\frac{1}{\eta}]) \;\;\;\;\;\; (\; s=\frac{e^{-x}}{t+1} \;)
\end{equation}
for {\bf A} and {\bf B} correspondingly. To determine the relevant
constants, $Y|_{x=0,\; t=0}=0$ and $\frac{\partial Y}{\partial
x}|_{x=0,\; t=0}=C$ are required here.

With the known magnetic fields, we can now solve the current and
the energy dissipation rate by Joule heat. Based on Eqs.(\ref{x9},
\ref{x10}), we achieve
\begin{equation}\label{y1}
j(x, t)=C \{\frac{1}{\sqrt{t+1}} {\bf {}_1F_1} [\frac{5}{6},
\frac{3}{2}, -\frac{3 x^2}{4 \eta (t+1)}] -
\frac{5 x^2}{6 \eta (t+1)^{3/2}} {\bf {}_1F_1} [\frac{11}{6},
\frac{5}{2}, -\frac{3 x^2}{4 \eta (t+1)}] \}
\end{equation}
for solution {\bf A}, and
\begin{equation}\label{y2}
 j(x, t)=C \{ 1+ \frac{1}{\eta (t+1)} e^{-\frac{e^{-x}}{\eta (t+1)}-
 x}({\bf\Gamma}[0, -\frac{e^{-x}}{\eta (t+1)}]- {\bf
\Gamma}[0, -\frac{1}{\eta}]) \}
\end{equation}
for solution {\bf B}. Finally, according to
\begin{equation}\label{y3}
{\cal W}_{\eta}=<\eta j^2>=\eta \int j(x, t)^2 dV
\end{equation}
we can calculate the power provided by Ohmic dissipation.

The dependence of $B(x, t), j(x, t)$ on $(x, t)$ and ${\cal
W}_\eta$ on time $t$ are shown in Figs.(2-7).

It can be seen from the figures that, for solution {\bf A}, the
maximum of the magnetic field departs from the initial magnetic
neutral interface with time. So, the dissipation area will get
gradually wide. Following it, the current becomes smaller and
smaller. The biggest rate of the diffuse of the magnetic energy due
to the Joule heat takes place at the beginning. The solution {\bf
B} is interesting, its current is evidently a soliton. Hence, when
the wave-packet of the current is located in the domain that we
consider, the dissipation power keeps as a constant. But, once it
passes through the domain, the dissipation power will descend with
time.

\vspace{0.5cm}
\centerline{\bf 4. Conclusions and Discussions}

In conclusion, we have obtained two classes of full analytic
time-dependent solutions for magnetic annihilation by directly
solving the MHD equations.

The solutions which we get here have such a scaling property that
the evolution of the relevant physical observables are described by
the solutions of certain ordinary differential equations, with the
variables being some functions of $x$ dilated by time $t$. Besides,
in the case of asymmetric inflow, a soliton-like current can be
formed. Finally, the solutions are used to reveal the relevant
physical characteristics in the process of the magnetic
annihilation.

However, the time-dependent solutions obtained here are the
relative simple ones. Based on the method provided by Watson, Craig
and Henton etc. \cite{craig1995,craig1996,watson1996,watson1997},
we expect to construct the time-dependent reconnection solutions.
The work along this direction is under way.

\vspace{0.5cm}

\noindent {\bf Acknowledgment}: One of the authors (Y. Liu) would like
to thank Dr. J. Q. Li for helpful discussions and Prof. P. G.
Watson for great valuable comments. This work is partly supported
by Chinese Postdoctoral Foundation.

\newpage
\begin{figure}[htb]
\mbox{}
\vskip 7in\relax\noindent\hskip -1 in\relax
\includegraphics{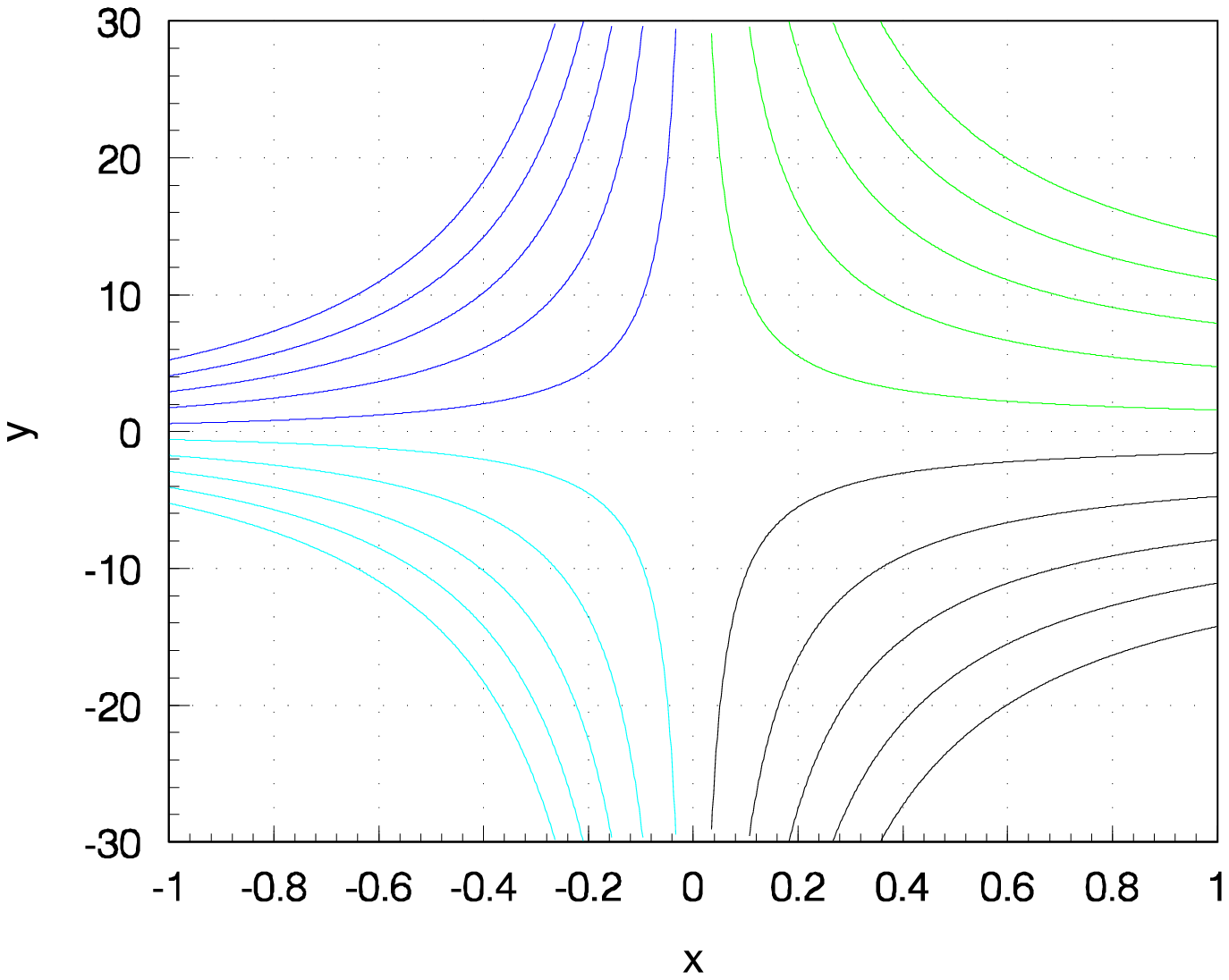}
\caption{Stream line for the asymmetric flow field. Where,
$\vec{v}(x, t)=\{\frac{e^{-x}-1}{t+1}, \;\; \frac{y e^{-x}}{t+1},
\;\; 0 \}.$}
\end{figure}

\begin{figure}[htb]
\mbox{}
\vskip 7in\relax\noindent\hskip -1 in\relax
\includegraphics{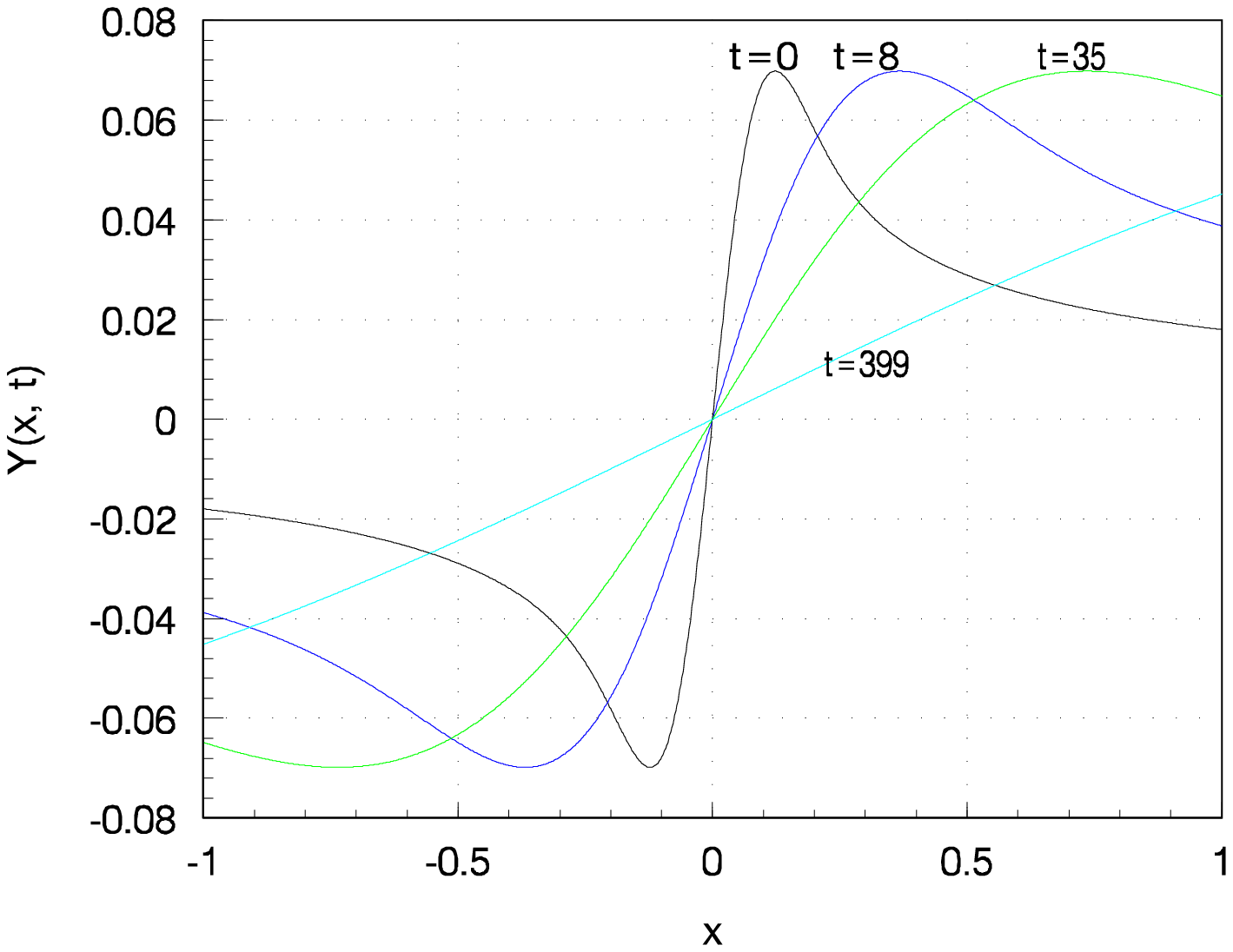}
\caption{ Magnetic field $Y(x, t)$ versus $x$ in different time $t$.
Where, $\eta=0.01, \;\; U(x,t)=-\frac{x}{t+1}$.}
\end{figure}

\begin{figure}[htb]
\mbox{}
\vskip 7in\relax\noindent\hskip -1 in\relax
\includegraphics{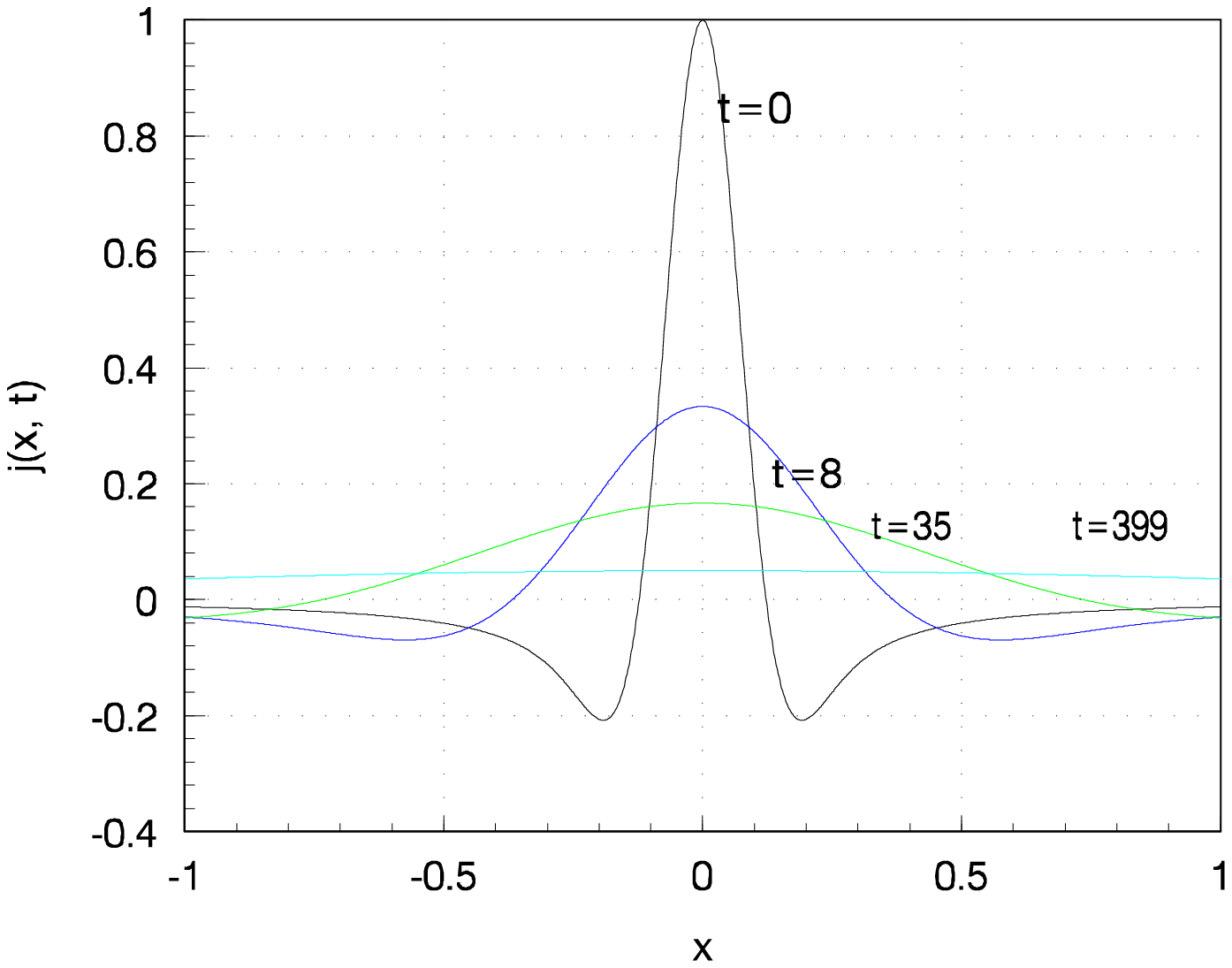}
\caption{Current $j(x, t)$ versus $x$ in different time $t$.
Where, $\eta=0.001, \;\; U(x,t)=-\frac{x}{t+1}$. }
\end{figure}

\begin{figure}[htb]
\mbox{}
\vskip 7in\relax\noindent\hskip -1 in\relax
\includegraphics{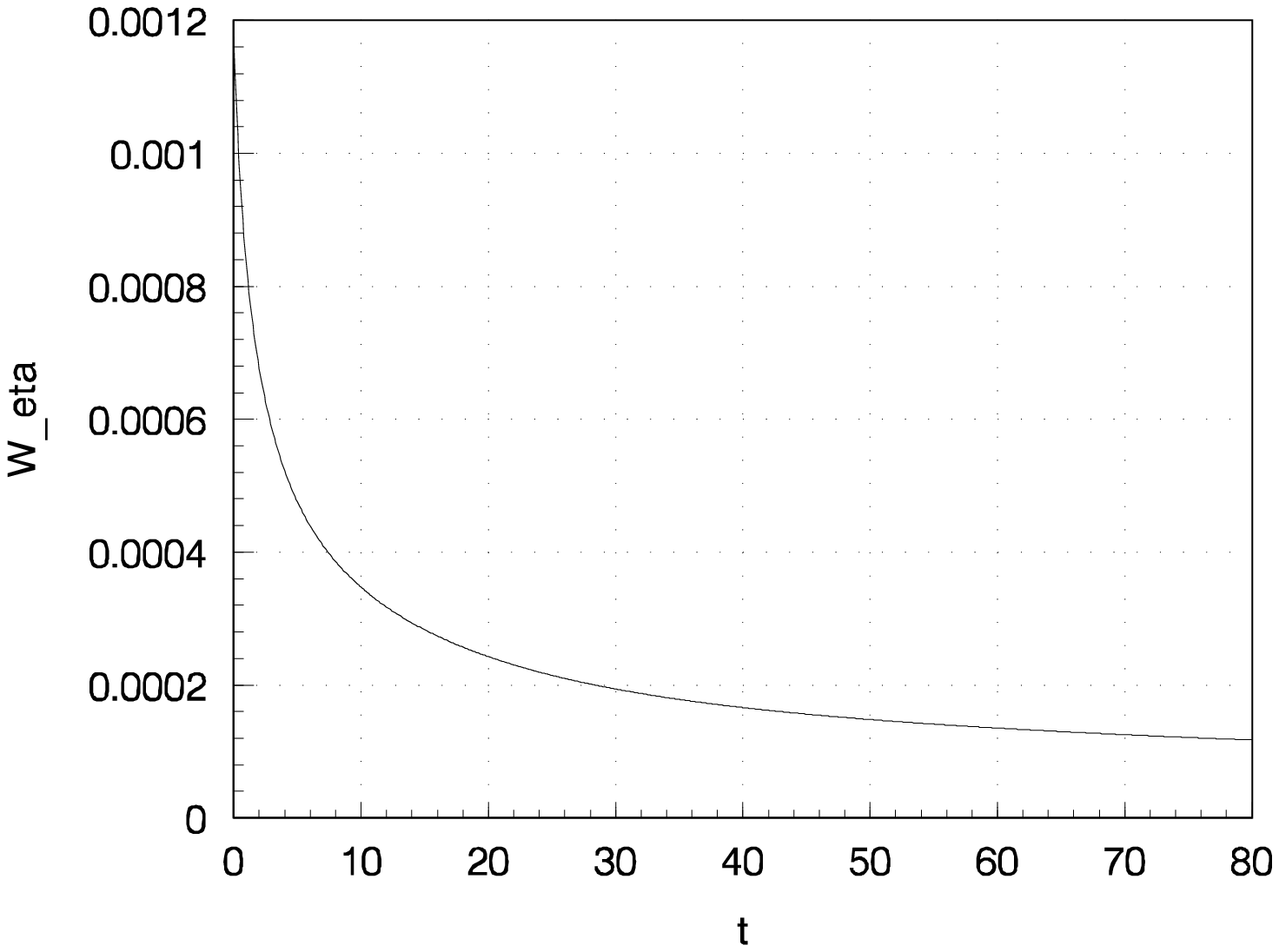}
\caption{Dissipation power of the magnetic energy versus time $t$ in the
calculated domain. Where, $\eta=0.01, \;\; -1 \leq x \leq 1,
\;\; U(x,t)=-\frac{x}{t+1}$. }
\end{figure}

\begin{figure}[htb]
\mbox{}
\vskip 7in\relax\noindent\hskip -1 in\relax
\includegraphics{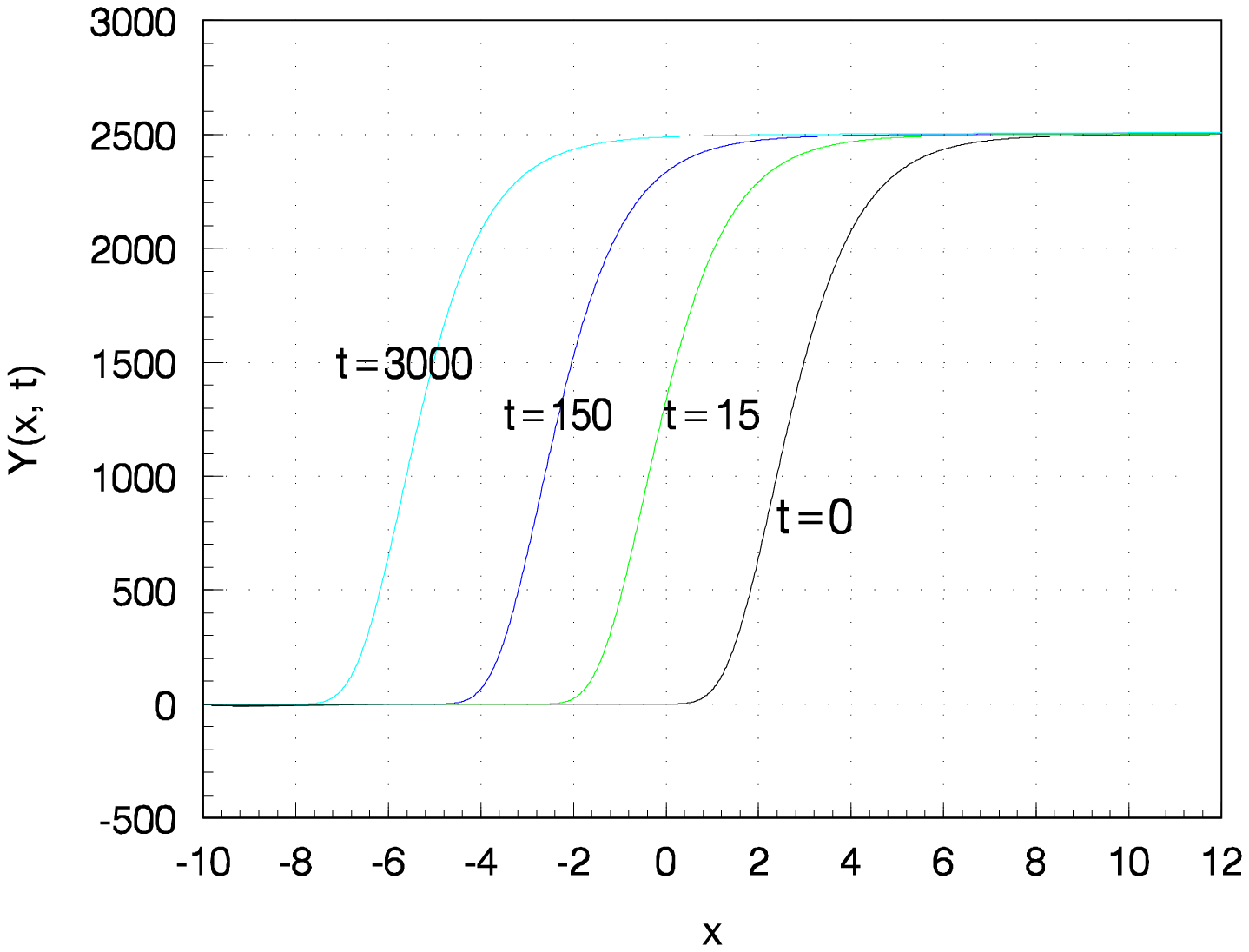}
\caption{Magnetic field $Y(x, t)$ versus $x$ in different time $t$.
Where, $\eta=0.1,\;\; U(x,t)=\frac{e^{-x}-1}{t+1}$. }
\end{figure}

\begin{figure}[htb]
\mbox{}
\vskip 7in\relax\noindent\hskip -1 in\relax
\includegraphics{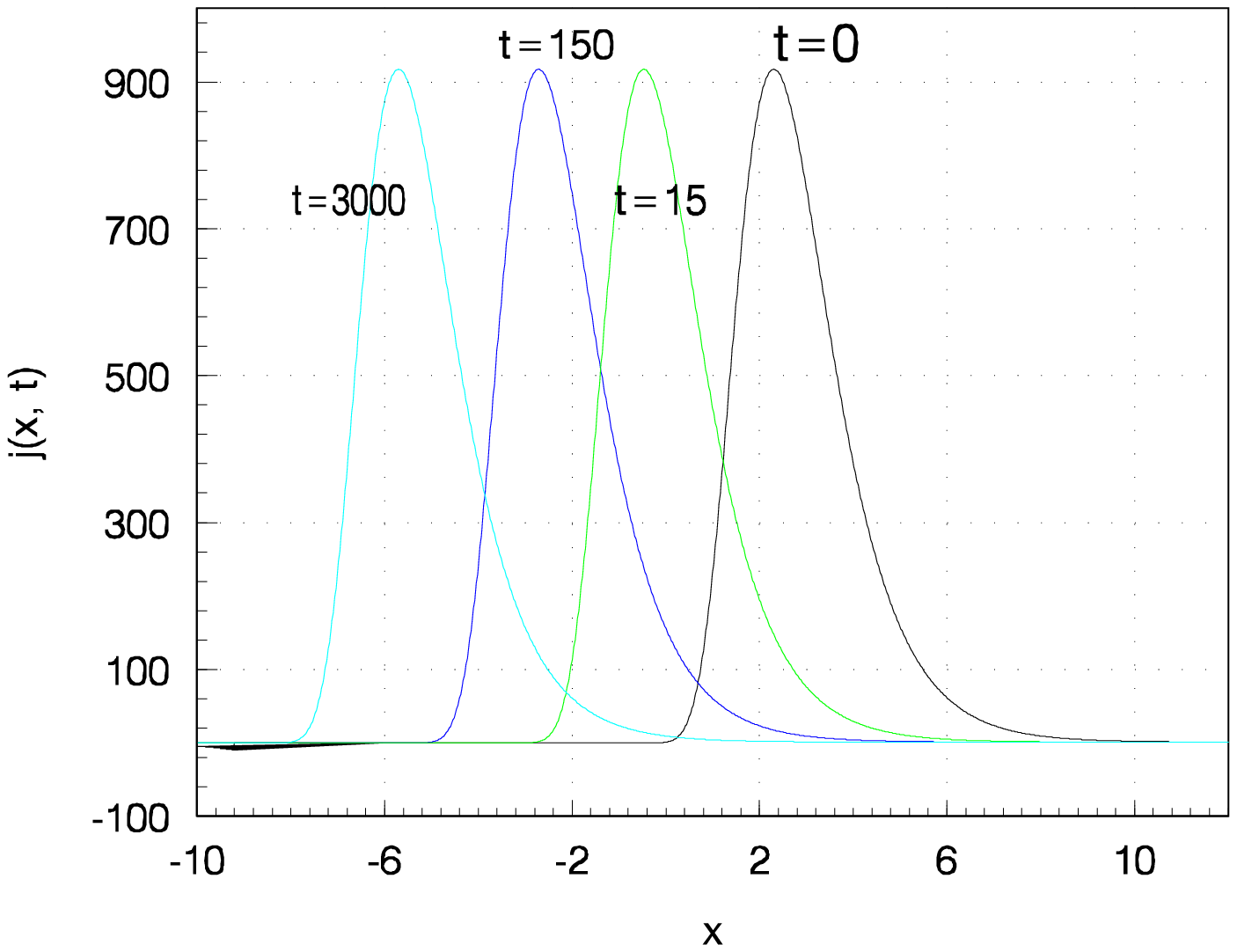}
\caption{Current $j(x, t)$ versus $x$ in different time $t$.
Where, $\eta=0.1,\;\; U(x,t)=\frac{e^{-x}-1}{t+1}$. }
\end{figure}

\begin{figure}[htb]
\mbox{}
\vskip 7in\relax\noindent\hskip -1 in\relax
\includegraphics{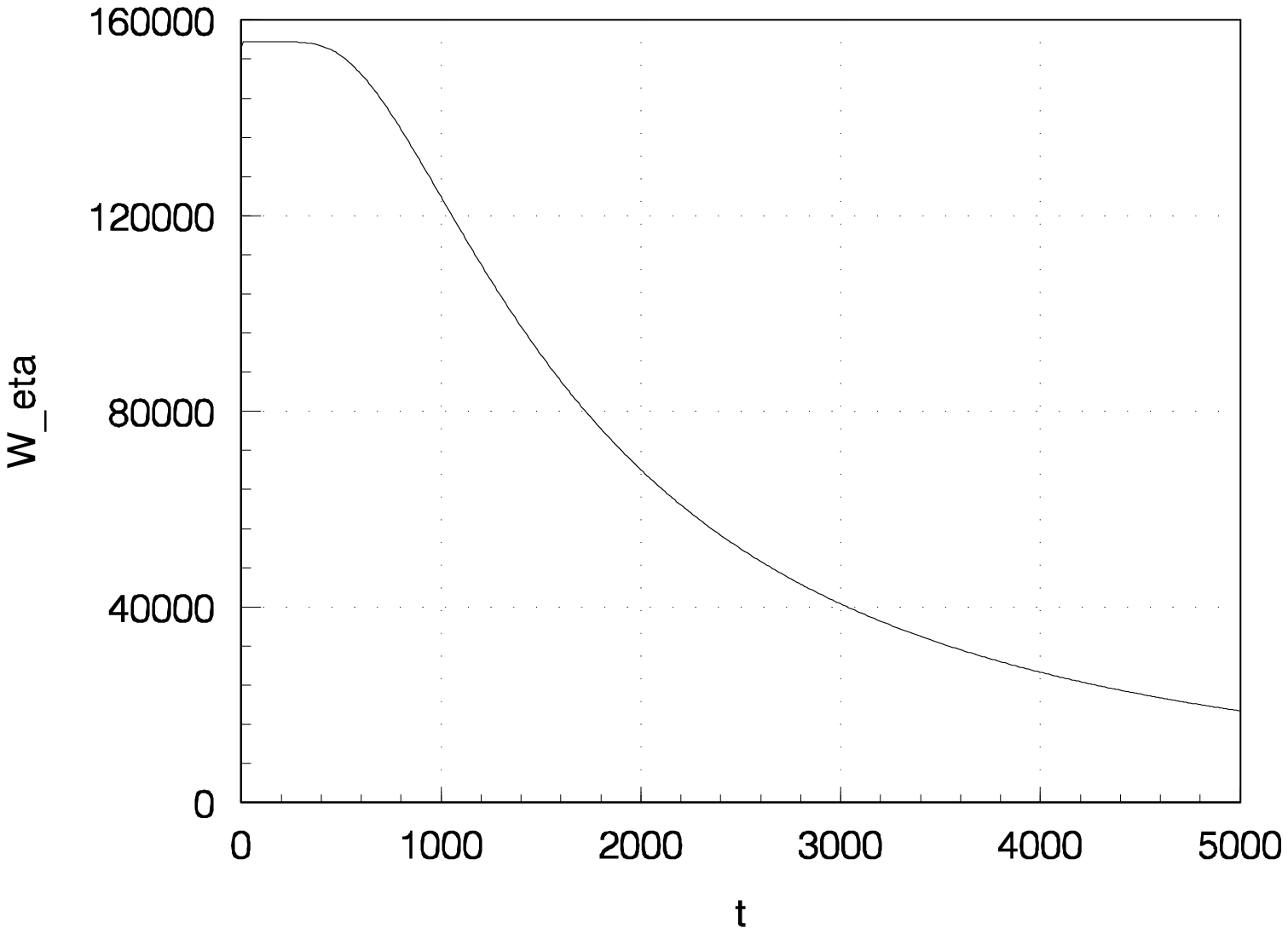}
\caption{Dissipation power of the magnetic energy versus time $t$
in the calculated domain versus time $t$. Where, $\eta=0.1, \;\; -5
\leq x \leq 5,\;\; U(x,t)=\frac{e^{-x}-1}{t+1}$. }
\end{figure}


\vspace{0.5cm}

\begin{thebibliography}{30}
\bibitem{sweet1958} P. A. Sweet, The production of high energy particles
in solar flares, {\it Nuovo Cimento, Suppl.}, {\bf 8}, 188-196, 1958.
\bibitem{parker1963} E. N. Parker, The solar flare phenomenon and the
theory of reconnection and annihilation of magnetic fields, {\it Astrophys.
J., Suppl., Ser. 77}, {\bf 8}, 177, 1963.
\bibitem{biskamp1994} D. Biskamp, Phys. Rev. Lett. {\bf 237}, 181(1994).
\bibitem{forbes1987} T. G. Forbes and E. R. Priest, Rev. Geophys. {\bf 25},
1583(1987).
\bibitem{parker1973} E. N. Parker, Comments on the reconnexion rate
of magnetic fields, {\it J. Plasma Phys.}, {\bf 9}, 49-63, 1973.
\bibitem{sonnerup1975} B. U. O. Sonnerup and E. R. Priest, Resistive MHD
stagnation-point flows at a current sheet, {\it J. Plasma Phys.}, {\bf
14}, 283-294, 1975.
\bibitem{phan1990} T. D. Phan and B. U. O. Sonnerup, MHD stagnation-point
flows at a current sheet including viscous and resistive effects:
General two-dimensional solutions, {\it J. Plasma Phys.}, {\bf 44},
525-546, 1990.
\bibitem{jardine1992} M. Jardine, H. R. Allen, R. E. Grundy and
E. R. Priest, A family of two-dimensional nonlinear solutions for
magnetic field annihilation, {\it J. Geophys. Res.}, {\bf 97}, 4199-4208,
1992.
\bibitem{heyn1993} M. F. Heyn and M. I. Pudovkin, A time-dependent model of
magnetic field annihilation, {\it J. Plasma Phys.}, {\bf 49}, 17-27, 1993.
\bibitem{heyn1996} M. F. Heyn and V. S. Semenov, Compressible
reconnection in compressible plasma, {\it Phys. Plasma}, {\bf 3},
2725-2741, 1996.
\bibitem{craig1995} I. J. D. Craig and S. M. Henton, Astrophys.
J. {\bf 450}, 280(1995).
\bibitem{craig1996} I. J. D. Craig and R. B. Fabling, Astrophys. J.
{\bf 462}, 969, 1996.
\bibitem{watson1996} P. G. Watson and I. J. D. Craig, Phys. Plasma {\bf 4},
101(1997).
\bibitem{watson1997} P. G. Watson and I. J. D. Craig, Phys. Plasma {\bf 4},
110(1997).
\bibitem{biernat1987} H. K. Biernat, M. F. Heyn and V. S. Semenov,
J. GeoPhys. Res.
{\bf 92}, 3392(1987).

\end{thebibliography}
\end{document}